\documentclass[aps,prl,showpacs,amsmath,amsfonts,superscriptaddress,twocolumn]{revtex4}

\usepackage{graphicx}
\usepackage{dcolumn}
\usepackage{bm}
\usepackage{color}

\begin{document}

\title{Incorporating Dynamic Mean-Field Theory into Diagrammatic Monte Carlo}

\author{Lode Pollet}
\affiliation{Theoretische Physik, ETH Zurich, 8093 Zurich, Switzerland}

\author{Nikolay V. Prokof'ev}
\affiliation{Department of Physics, University of Massachusetts,
Amherst, MA 01003, USA}
\affiliation{Russian Research Center ``Kurchatov Institute'',
123182 Moscow, Russia}

\author{Boris V. Svistunov}
\affiliation{Department of Physics, University of Massachusetts,
Amherst, MA 01003, USA}
\affiliation{Russian Research Center ``Kurchatov Institute'',
123182 Moscow, Russia}

\begin{abstract}
The bold diagrammatic Monte Carlo (BDMC) method performs an unbiased sampling
of Feynman's diagrammatic series using skeleton diagrams.
For lattice models the efficiency of BDMC can be dramatically
improved by incorporating dynamic mean-field theory solutions into
renormalized propagators. From the DMFT perspective, combining it with BDCM leads to an unbiased method with well-defined accuracy. We illustrate the power of this approach by computing the single-particle propagator (and thus the density of states) in the non-perturbative regime of the Anderson localization problem, where a gain of the order of $10^4$ is achieved with respect to conventional BDMC
in terms of convergence to the exact answer.
\end{abstract}



\date{\today}
\maketitle

A skeleton diagrammatic series is nothing but
Feynman's diagrammatic expansion in terms of  `dressed', or `bold-line',
propagators, interaction lines, and vertices,
which account for the summation of certain subclasses of diagrams.
Its power lies in the fact that,
even when truncated to the lowest orders, it often captures the basic physics of strongly correlated systems and yields quantitatively accurate answers.
Among its numerous successful examples we mention screening effects,
self-consistent Hartree-Fock schemes, the GW-approximation for simple metals,
Bogoliubov and Gor'kov-Nambu equations, etc. Often, as, e.g., in case
of Kohn-Sham orbitals in density functional theory, the diagrammatic
structure is hidden in a set of integral equations,
whose implementation has been improved to perfection.
Physically, the lowest-order skeleton graphs embody the idea of
incorporating some `mean-field' theory self-consistently.  \\

The notorious shortcoming of self-consistent treatments based on the lowest-order
diagrams is lack of accuracy and control: the error due to truncation can be
established only by reliably calculating contributions
of higher-order diagrams, which in the typical case of optimized codes
solving a set of self-consistent integral equations is nearly impossible
(in the absence of small parameters order of magnitude estimates
are essentially meaningless).
The recently developed  bold diagrammatic Monte Carlo (BDMC) method \cite{bdmc}
allows one to sample skeleton Feynman's expansions far beyond the mean-field level.
Given that even the diagrammatic Monte Carlo method based on bare propagators can produce
very accurate results for correlated systems (say, for the repulsive
fermionic Hubbard model \cite{f_Hubbard}), BDMC emerges as a powerful
generic field-theoretical method. It has been successfully applied to the
fermi-polaron problem \cite{bdmc}, and, very recently, to the problem of
equation of state in a system of resonant fermions \cite{BCS-BEC}.
The above examples deal with continuous-space problems,
but it is natural to expect that working with the skeleton series will
bring significant advantages to lattice models as well.

\begin{figure}[h]
 \includegraphics[angle=0, width=1.0\columnwidth]{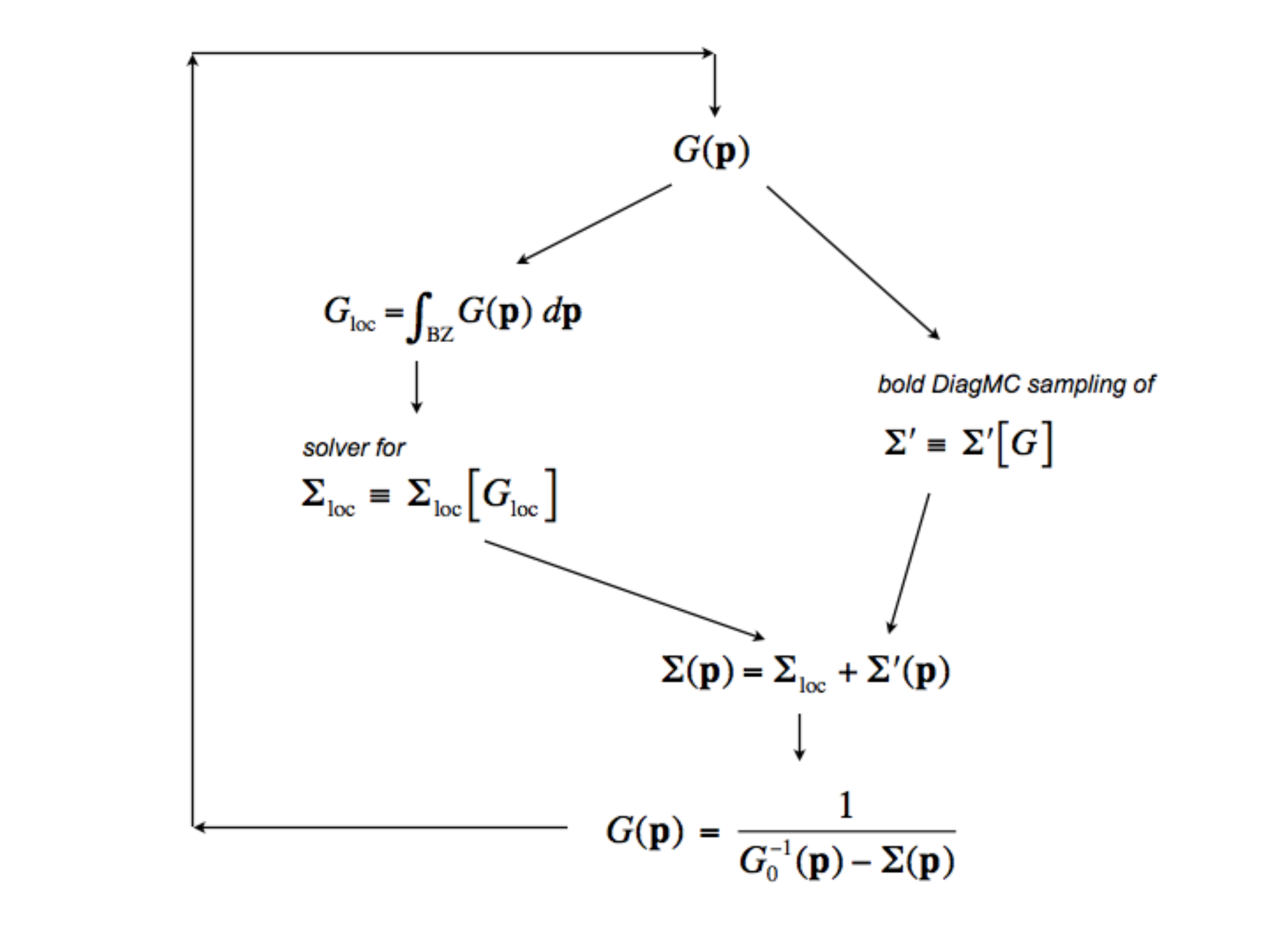}
 \caption{Schematic representation of the BDMC+DMFT protocol (see text for the details). }
 \label{fig:schema}
\end{figure}

In this Letter, we show that, in addition to simply going from bare to
skeleton expansion, a  dramatic increase in performance can be reached
by employing an exact series re-summation procedure which accounts for
the summation of all local contributions to the self-energy.
This approach amounts to embedding the dynamic mean-field theory
(DMFT)\cite{DMFT1} solution into an exact diagrammatic method
and avoids, in particular, any double counting or other uncontrollable errors.
The gain in efficiency comes from two related observations:
an impressive success of DMFT applications \cite{DMFT1,DMFT2},
and the fact that summation of local contributions
can be done separately by a variety of highly efficient methods.
The BDMC+DMFT approach thus involves two distinct but cross-linked
numerical processes: (i) a problem-specific solver of the DMFT-type problem
(to be referred to as `impurity solver', in accordance with
terminology accepted in literature), and (ii) a generic BDMC scheme
simulating skeleton diagrams which cannot be reduced to the purely local
ones. The protocol is illustrated in Fig.~\ref{fig:schema}. \\

Below we start with the precise formulation of the combined scheme, and then
proceed with its implementation for finding a disorder-averaged single-particle
propagator (and thus the density of states) in the non-perturbative
regime of the Anderson localization problem which is well suited for illustrating
the idea because
the efficiency gained by incorporating DMFT solutions within the
BDMC is about $10^4$.
In general, the gain will be problem and parameter specific, and will
also depend on the efficiency of the impurity solver.
[We stress that our goal in this article is to explain the new method and illustrate
its implementation, not to solve the localization problem in its full complexity.] \\

{\it Formalism.} -- The protocol of reformulating skeleton series to account for all local
contributions to self-energy is conceptually straightforward.
The Dyson equation relates the Green's function, $G$,
to the self-energy, $\Sigma$ (for clarity, we suppress below the frequency
variable):
\begin{equation}
G({\mathbf p}) \, = \, \frac{1}{G_0^{-1}({\mathbf p})- \Sigma({\mathbf p})} \, ,
\label{protocol1}
\end{equation}
with $G_0$ standing for the non-perturbed Green's function.
The local propagator $G_{\rm loc}$ is defined by
integrating over the Brillouin zone `BZ'
\begin{equation}
G_{\rm loc} \, = \, \int_{\rm BZ} \,
G({\mathbf p}) \, \frac{d {\mathbf p}}{(2\pi)^d} \; .
\label{protocol2}
\end{equation}
We now separate contributions to the self-energy into two parts
\begin{equation}
\Sigma({\mathbf p}) = \Sigma_{\rm loc} + \Sigma'({\mathbf p})  \;,
\label{protocol3}
\end{equation}
where $\Sigma_{\rm loc}$ is given by irreducible skeleton diagrams
which involve {\it exclusively} $G_{\rm loc}$ propagators. In other words,
this local propagator has only purely momentum independent building blocks,
while all the rest is put in $\Sigma^{\prime}$. \\

Numerically, one calculates the self-energy using current knowledge of
the Green's function and then uses it to permanently improve the knowledge
of $G$ within the self-consistent process. This involves two steps.
First, the current knowledge of $G_{\rm loc}$ serves as an input for
the calculation of $\Sigma_{\rm loc}\equiv \Sigma_{\rm loc}[G_{\rm loc}]$
achieved by the impurity solver, and $G({\mathbf p})$ is used for the BDMC
simulation of the remaining skeleton graphs.
Second, self-energies  $\Sigma_{\rm loc}$ and $\Sigma' $ are combined into the total self-energy, Eq.~(\ref{protocol3}), which is then used to
find the updated $G$ by Eq.~(\ref{protocol1}). This is illustrated in
Fig.~\ref{fig:schema}. \\

Technically, the crucial advantage of separating local contributions to the self-energy is that the corresponding momentum independent problem
admits a variety of techniques for solving it very efficiently~\cite{impurity_solvers}. Treating the local physics non-perturbatively is very appealing from the physical viewpoint. In typical problems such as the Hubbard model, the diagrammatic technique expands around the non-interacting limit which is dominated by large hopping processes. The competing phase with large on-site interactions tends on the contrary to localize the particles. 
Hence, building diagrams on top of the solution capturing essential physics of the competing phase may be better suited for describing the difficult intermediate regime as well. Local physics is also dominant 
at high temperatures which can easily be understood in terms of Feynman's path integrals. \\

From Eqs.~(\ref{protocol1})-(\ref{protocol2}) it is explicitly seen that
BDMC+DMFT process  builds an {\it  exact} solution of the problem
on top of the DMFT answer, which is crucial not only for improving the quality of the final result but also for reliable estimates of corrections to mean-field results. \\

One of the solvers for obtaining $\Sigma_{\rm loc}$ in terms of $G_{\rm loc}$
widely used in the standard DMFT approach is based on an {\it implicit} formulation of the problem in terms of the single-site (or impurity) effective action with a certain auxiliary (to be determined) `bare'' propagator $\tilde{g}_0$. The advantage of this formulation is in the
flexibility of designing efficient tools (impurity solvers) \cite{impurity_solvers}
for obtaining the $G_{\rm loc}[\tilde{g}_0]$ relation;
the local self-energy readily follows from $\Sigma_{\rm loc}= \tilde{g}_0^{-1} -G_{\rm loc}^{-1}$.
Iterations leading to the self-consistent solution consist of plugging
the thus obtained self-energy in Eq.~(\ref{protocol2}) to redefine the auxiliary
propagator by $\tilde{g}_0^{-1} = G_{\rm loc}^{-1}[\Sigma]+\Sigma_{\rm loc}$.
Solvers based on the effective action approach play a crucial part when the diagrammatic
expansion of $\Sigma_{\rm loc}[G_{\rm loc}]$ cannot be used because of  technical or convergence problems. \\

{\it Illustration.} -- We illustrate the introduced concepts for Anderson's model of
particle localization on a disordered three-dimensional cubic lattice.
We consider delta-correlated gaussian disorder in the chemical potential,
for which the standard diagrammatic technique can be formulated~\cite{AGD}.
The Hamiltonian, in standard lattice notation, reads
\begin{equation}
H = -J \sum_{ \langle i,j \rangle}\,  \hat{c}_i^{\dagger} \hat{c}_j \, + \, \sum_i \, \epsilon_i \, \hat{n}_i\, .
\end{equation}
The random on-site potential  $\epsilon_i$ is distributed with the gaussian probability density 
\begin{equation}
P(\epsilon) = \frac{e^{-\epsilon^2/2V^2}}{\sqrt{2\pi V^2}} \, ,
\end{equation}
the dispersion $V$ characterizing the strength of the disorder. We choose $J=1$ as our unit.
 We work in the real-time representation
where the Green function is defined as $G({\bf r}, t'\! -\!t) = -i \, \langle \,  {\mathcal {T}} c(0, t)\,  c^{\dagger}({\bf r}, t')\,  \rangle$.
We took a lattice of size $12 \times 12 \times 12$. Just like in conventional DMFT, larger lattices pose no problem at all; in fact, larger lattices would suppress revivals and make hence the simulations easier. 
 The (local) density of states is given by the imaginary part of its Fourier transform  for ${\bf r} =0$, ${\rm DOS} (\omega) = - \pi^{-1} {\rm Im} G( {\bf r}=0, \omega)$,  which can  be compared with the  exact diagonalizaton results of Refs.~\cite{Kravtsov, Wortis}. \\

\begin{figure}[h]
 \includegraphics[angle=0, width=0.9\columnwidth]{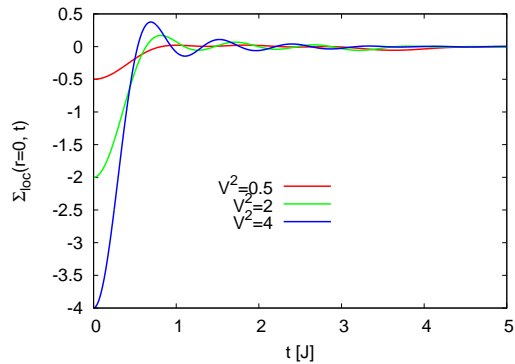}
 \caption{(Color online) Local self-energy calculated using local 
 propagators ($\Sigma^{\prime}=0$) for disorder strengths $V=1/\sqrt{2}, ~\sqrt{2},~4$ on a lattice of size $12 \times 12 \times 12$.
 }
 \label{fig:self}
\end{figure}

Evaluating the sum of all skeleton diagrams involviong local propagators only (i.e., the DMFT part~\cite{Byczuk10}) simplifies for Anderson's localization since disorder lines have no time dependence.
For a single-site problem, one does not  even need to expand the gaussian exponential into the diagrammatic series, because averaging the Green's function---in the frequency representation,
the former is immediately found to be  equal to $1/[1/\tilde{g}_0 (\omega) - i \epsilon  ]$---over the disorder amounts to performing 
a simple one-dimensional integral:
\begin{equation}
G_{\rm loc}(\omega) =\frac{ \tilde{g}_0 (\omega)}{\sqrt{2\pi V^2}} \int \frac{e^{-\epsilon^2/2V^2}}{ 1 - i \epsilon \tilde{g}_0 (\omega) } \, d\epsilon \;.
\label{eq:integral_representation}
\end{equation}
The local self-energy then follows from $\Sigma_{\rm loc}(\omega)=\tilde{g}_0^{-1} (\omega) - G_{\rm loc}^{-1}(\omega)$ which accounts for the implicit (parametric) complex-number
relation $\Sigma_{\rm loc}[G_{\rm loc}]$, i.e. the goal is achieved by the semi-analytic exact solution. In practice this is done by a parametrization
of the above integral equation through $\tilde{g}_0[G_{\rm loc}]$ (inversion),
and iterating until self-consistency is reached.
This works fine here only because the interaction lines carry no time dependence.
In Fig.~\ref{fig:self} we show for various disorder strengths the local self-energy
obtained for $\Sigma'=0$ after convergence, i.e. the answer as predicted by the
conventional DMFT approach.\\

\begin{figure}[t]
  \includegraphics[angle=0, width=0.9\columnwidth]{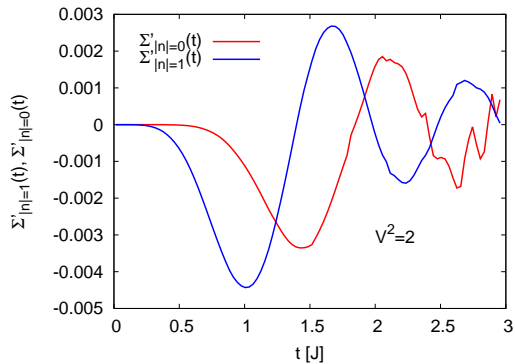}
  \caption{(Color online) Correction to local self-energy for $V = \sqrt{2}$  on a lattice of size $12 \times 12 \times 12$:
$\Sigma'_{\vert {\mathbf n}\vert =0}(t)$  (red line) and 
$\Sigma'_{\vert {\mathbf n}\vert =1}(t)$  (blue line). The noise in the curve is indicative of the error bars. The sign problem and the high expansion orders
put a limit on the accessible times.  }
  \label{fig:selfprime}
\end{figure}

The full calculation involves Monte Carlo sampling of all skeleton
diagrams except those contributing to $\Sigma_{\rm loc}$
(which would otherwise consume about $90 \%$ of the simulation time already for $V=\sqrt{2}$). In the real-space representation this means that only skeleton
graphs which contain at least two vertices with different site indices are
accounted for in $\Sigma'$. The simulation itself was done using standard BDMC
rules with the self-consistency loop implemented exactly as described in
the introductory part of this Letter. It turns out that the diagrammatic series
for Anderson's localization problem constitutes the `worst case scenario' in terms
of convergence properties. Although for any finite time $t$ the series are convergent
(allowing us to use Dyson's equation and Eq.~(\ref{eq:integral_representation})), the required expansion order increases dramatically with the time $t$.
Realistically, we were able to deal with skeleton graphs up to order
$n_{\rm max}\sim 50$ which was limiting the accessible times in the simulation of
$\Sigma'$. We observe that the values of $\Sigma'_{{\mathbf r} -{\mathbf r}'}\equiv \Sigma'_{\mathbf n}$ turn out to be extremely small, about two orders of
magnitude smaller than $\Sigma_{\rm loc}$ even in the intermediate
coupling regime $V=\sqrt{2}$, see Fig.~\ref{fig:selfprime}. Since the complexity, 
and hence the {\it relative} error-bar, of the BDMC simulation is roughly the 
same for simulating $\Sigma$ or
$\Sigma'$, we conclude that the BDMC+DMFT scheme produces results
which are two orders of magnitude (or a speedup of $\sim 10^4$ in CPU time) more accurate for the same simulation time
in the region of parameter space where the series converges and error bars are under
control. This constitutes
the proof of principle for the proposed scheme. Final results for the density
of states are indistinguishable from the exact diagonalization data. 
 
{\it Outlook and Conclusions.} -- We have introduced an approach that uses DMFT as an integral part of performing simulations of skeleton graphs in 
strongly interacting systems. It combines the power of solving impurity problems 
efficiently with the diagrammatic formalism that is unbiased and exact.
Given the already good agreement between DMFT and diagrammatic Monte Carlo  based on bare propagators 
for the Hubbard model at $U/J=4$~\cite{Kozik10}, we expect the present formalism to 
bring radical speed up and accuracy to studies of the Hubbard model at larger 
values of $U$ and lower temperatures.  \\

We would  also like to mention several generalizations of the simplest scheme
introduced above. 
To begin with, the definition of momentum-independent
propagator allows the use of an arbitrary function $f({\mathbf p})$ in the definition of $G_{\rm loc}$ such that
$G_{\rm loc} =  \int_{\rm BZ} \, G({\mathbf p}) f({\mathbf p}) \, d {\mathbf p}/ (2\pi)^d$. The rest of the scheme remains intact: as before diagrams 
containing {\it exclusively} $G_{\rm loc}$ propagators are all summed up in the
local self-energy while $\Sigma'$ contains at least one line which is based on
$G({\mathbf p})-G_{\rm loc}$. The freedom of choosing $f({\mathbf p})$ different from
a constant may be used to optimize the subtraction of leading terms. \\

In the generic many-body skeleton diagram, any renormalized line whether it is the single-particle propagator  $G({\mathbf p})$, the interaction line $W({\mathbf q})$, or
the two-particle propagator $\Gamma $, can be split into 
momentum-independent and momentum-dependent parts (with the same freedom of 
defining the local part as described in the previous paragraph). Next, all
diagrams based {\it exclusively} on momentum-independent lines can be dealt with
using impurity solvers with BDMC accounting for the remaining graphs. Since 
the summation of certain geometric series such as ladder or screening diagrams can
be done analytically to set up the original diagrammatic space, one can go even further
beyond the purely local physics by doing so. \\

Our final remark is that nothing prevents one from extending the idea of subtracting 
diagrams with momentum-independent lines (and compensating them separately by impurity solvers) to subtracting diagrams with specific momentum-dependence and structure,
(and compensating them by impurity solvers dealing with a few sites, 
similar to the ideas behind cluster-DMFT schemes). The diagrams to be summed up
by the impurity solver are those with the connections of a compact cluster 
of sites. Similar extensions for real-space clusters are also possible. \\

This work was supported by the National Science Foundation
grant PHY-1005543, the Swiss National Science Foundation under grant PZ00P2-131892/1,  and by a grant from the Army Research Office
with funding from the DARPA OLE program. We thank the Aspen Center of Physics, KITP Santa Barbara, and Casa F{\'i}sica at UMass for hospitality.
Simulations were performed on the Brutus cluster at ETH Zurich and CM cluster at UMass, Amherst.

\end{document}